\documentclass[pre,twocolumn,floatfix,aps,showpacs]{revtex4}
\usepackage{amssymb}
\usepackage{graphicx}
\usepackage{epsf}
\usepackage{amsmath}
\usepackage{bm} %bold math
\usepackage{pspicture}
\usepackage{flafter}
\usepackage{float}
\begin{document}

\title{Continuous and discontinuous phase transitions and partial
synchronization in stochastic three-state oscillators}
\author{Kevin Wood$^{1,2}$, C. Van den Broeck$^{3}$, R. Kawai$^{4}$,
and Katja Lindenberg$^{1}$}
\affiliation{
$^{(1)}$Department of Chemistry and Biochemistry and Institute for
Nonlinear Science, and $^{(2)}$ Department of Physics,
University of California San Diego, 9500 Gilman Drive, 
La Jolla, CA 92093-0340, USA\\
$^{(3)}$Hasselt University, Diepenbeek, B-3590 Belgium\\
$^{(4)}$ Department of Physics, University of Alabama at Birmingham,
Birmingham, AL 35294 USA
}
\date{\today}

\begin{abstract}
We investigate both continuous (second-order) and discontinuous
(first-order) transitions to
macroscopic synchronization within a single class of discrete, stochastic
(globally) phase-coupled oscillators.  We provide analytical
and numerical evidence that the continuity of the transition
depends on the coupling coefficients and, in some nonuniform
populations, on the degree of quenched disorder.
Hence, in a relatively simple setting this class of models exhibits
the qualitative
behaviors characteristic of a variety of considerably more complicated
models.  In addition, we study the microscopic basis of synchronization above
threshold and detail the counterintuitive subtleties relating
measurements
of time averaged frequencies and mean field oscillations.  Most notably,
we observe a
state of suprathreshold partial synchronization in which time-averaged
frequency measurements from individual oscillators do not correspond to
the frequency of macroscopic oscillations observed in the population.

\end{abstract}
\pacs{64.60.Ht, 05.45.Xt, 89.75.-k}

\maketitle

\section{Introduction}
A great number of physical systems consist of individual entities
with periodic, or nearly periodic, dynamics. Ranging from collections
of chemical consitutents to groups of social entities -- for example,
applauding individuals whose clapping is repetitive -- these systems serve
as a battleground of sorts for the competition between the dynamics
of individual constituents and the large scale cooperation favored in
many cases by the nature of their mutual interactions.  Owing to the
ubiquity and certainly, in part, to the dramatic nature of the
emergent synchronized behavior in such naturally oscillating settings,
the subject has been intensely studied in the physics literature for
several decades, with the Kuramoto oscillator and its kin serving as
prototypical models on which many studies are
based~\cite{strogatz,winfree,kuramoto,strogatz2}.

Recently, a simple class of models of macroscopic synchronization have
provided a number of additional insights into the large-scale phenomena
ocurring in noisy discrete coupled oscillators, including detailed
characterizations of both the universal critical behavior of the
continous phase transition~\cite{threestate1,threestate2} as well as the
effects of spatial disorder in such populations~\cite{threestate3}. 
While retaining many qualitative characteristics of more complex models,
the discrete oscillators remain sufficiently simple to provide results
unattainable in most of the paradigmatic settings.  

In this paper, we generalize our class of stochastic, discrete
oscillator models and detail its use in a variety of new contexts. 
By generalizing the form of the inter-oscillator coupling, we show that
our class of mean field models encompasses oscillators which can undergo
either supercritical or subcritical Hopf bifurcations, depending on
the microscopic specifics of the coupling.  In addition, we study
dichotomously disordered populations of oscillators and show that the
bifurcation can be either supercritical or subcritical depending on the
degree of disorder in the population.  Such behaviors are reminiscent of
a number of significantly more complex oscillator
models~\cite{gianuzzi07,filatrella07,tanaka,acebron,choi,pazo,hong0},
including Daido's generalized
Kuramoto oscillators~\cite{daido}, where either disoder or microscopic
coupling specifics can alter the nature (continuous or discontinuous)
of the transition.  However, our model provides a far simpler setting for
observing both continous and discontinous transitions to synchronization.  

In addition, we study the microscopic underpinnings of synchronization
above threshold.  In particular, we look at time-averaged frequency and
its relationship to phase synchronization above threshold (which turns
out to be interestingly counterintuitive).  Again, we
do this for a specific model from our general class (for both single
and dichotomously disordered populations), but we expect the results
to hold for our entire class of models undergoing a Hopf bifurcation. 
This is somewhat similar to the partial synchronization seen in other
models~\cite{rosenblum07}, but again, our model is simpler and, perhaps,
more transparent. 

In Sec.~\ref{sec2} we present our model and highlight its essential
parameters.  Here we note that for systems of
oscillators with identical transition rates between states the
control parameter for the phase transition is the coupling strength
among oscillators; when the array includes oscillators with 
different transition rates, the degree of disorder is also a control
parameter.  In Sec.~\ref{TransitionSec}
we show that depending on the values of microscopic parameters,
this model can exhibit both subcritical (or first-order)
and supercritical (or second-order) phase transitions, as a function of
the coupling strength and also as a function of the degree of disorder.
Section~\ref{underpinnings} deals with the microscopic underpinnings of
the synchronization phenomenon and the connection between phase
synchronization and frequency entrainment in our system.
Section~\ref{discussion} presents a summary and further discussion of
our results.

\section{The Model}
\label{sec2}
\begin{figure}[b]
\begin{center}
\includegraphics[width=3.0 cm]{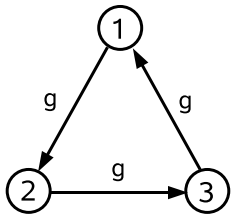}
\caption{Three-state unit with transition rates $g$.}
\label{fig1}
\end{center}
\end{figure}

In this section we present our model in some detail, repeating some of our
early presentations~\cite{threestate1,threestate2,threestate3}
because of important (albeit simple) generalizations of the model.
We begin by considering a stochastic three-state model governed by
transition rates $g$ (see Fig.~\ref{fig1}), where each state may be
interpreted as a discrete phase~\cite{threestate1,threestate2,threestate3}. 
Because the transitions among states are unidirectional and do not conform
to deterministic rate laws, the model retains a qualitative link with a
noisy phase oscillator.  
The linear evolution equation of a single oscillator is
$\partial \boldsymbol{P}(t)/\partial t = M \boldsymbol{P}(t)$,
where the components $P_i(t)$ of the
column vector $P(t)= (P_1(t)~P_2(t)~P_3(t))^T$ ($T$ denotes the
transpose) are the probabilities of being in states $i$ at time $t$,
where $i=1,2,3$ and
\begin{equation}
M = \begin{pmatrix} -g & 0 & g \\ g & -g & 0 \\
0 & g & -g \end{pmatrix}.  
\label{Mmat}
\end{equation}
The system reaches a steady state for $P_1^*=P_2^*=P_3^*=1/3$.
The oscillator's periodicity, as contained in the timescale of
the cycle $i=1\rightarrow 2 \rightarrow 3 \rightarrow 1...$, 
is determined by $g$; that is, the time
evolution of our simple model qualitatively resembles that of the
discretized phase of a generic noisy oscillator with the
intrinsic eigenfrequency set by the value of $g$.

To study interacting arrays of these oscillators, we couple individual
units by allowing the transition rates of each unit to depend on
the states of the units to which it is connected.  Specifically,
for $N$ identical units we choose the transition rate of a unit $\nu$
from state $i$ to state $i+1$ as 
\begin{equation}
g_{i} = g \exp\left[{\frac{a( U N_{i+1}+ V N_{i-1} + W N_{i})}{n}}\right],
\label{gnu}
\end{equation}
where $i=1,2,3$ and $i+1=1$ when $i=3$, $a$ is the coupling parameter,
$g$ is the transition rate parameter, $n$ is the number of oscillators to
which unit $\nu$ is coupled, and $N_k$ is the number of units among the
$n$ that are in state $k$.  We introduce the real constants $U$, $V$,
and $W$ to encompass in a general way our two previous coupling
functions~\cite{threestate1,threestate2,threestate3}.
Each unit may thus transition to the state ahead or remain in its
current state, and the propensity for such a change depends on the states
of its nearest neighbors. 
In our earlier works we considered the globally coupled system, $n=N-1$,
and also nearest neighbor coupling in square, cubic, or hypercubic
arrays, $n=2d$ ($d=$ dimensionality).  Here we focus on globally
coupled arrays.

For a population of $N\to\infty$
identical units in the mean field (globally coupled)
version of this model 
we can replace $N_k/N$ with the probability $P_k$, thereby arriving
at a nonlinear equation for the mean field probability,
$\partial \boldsymbol{P}(t)/\partial t =
M[\boldsymbol{P}(t)]\boldsymbol{P}(t)$,  with
\begin{equation}
M[\boldsymbol{P}(t)]
= \begin{pmatrix} -g_{1} & 0 & g_{3} \\ g_{1} & -g_{2}
& 0 \\ 0 & g_{2} & -g_{3} \end{pmatrix}.  
\label{Mmatmn}
\end{equation}
Normalization allows us to eliminate $P_3(t)$ and obtain a closed set
of equations for $P_1(t)$ and $P_2(t)$.  We can then linearize about
the fixed point $(P_1^*,P_2^*)=(1/3,1/3)$, yielding a Jacobian
$A(a,g,U,V,W)$ with a set of complex
conjugate eigenvalues which determine the stability of this
asynchronous state.  Specifically, we find that
\begin{equation}
\begin{aligned}
\lambda_\pm &= C \big(-9 + 3 a \Delta_{UW} \\
& \pm i \sqrt{3}(3 + a(U+W-2V) ) \big), 
\end{aligned}
\end{equation}
where $C \equiv g e^{a(U+V+W)/3}/6 $ is a nonzero constant for all
finite $U$, $V$, and $W$ and we introduce the abbreviation
$\Delta_{mn} \equiv m-n$.  The
eigenvalues cross the imaginary axis at $a_c=3/\Delta_{UW}$, yielding
\begin{equation}
\lambda_{\pm}^*= \pm i \omega(U,V,W)
\end{equation}
with
\begin{equation}
\label{omegaeqn}
\omega(U,V,W) \equiv g \sqrt{3}e^{a(U+V+W)/\Delta_{UW}}
\frac{\Delta_{UV}}{\Delta_{UW}}.
\end{equation}
For $\Delta_{UW} \neq 0$ and $\omega(U,V,W) \neq 0$ (that is,
$\Delta_{UV} \neq 0$), $a_c$ represents a Hopf bifurcation point,
indicating the emergence of macroscopic oscillations indicative of
synchronization.  Furthermore, we require that $\Delta_{UW} > 0$ to
ensure the bifurcation happens at a positive value of $a$.  We note
that in previous studies we have used $(U,V,W)=(1,-1,0)$, yielding
$a_c=3$ and $\omega= 2 g \sqrt{3}$~\cite{threestate3},
and $(U,V,W)=(1,0,-1)$, yielding $a_c=1.5$ and
$\omega=g \sqrt{3}/2$~\cite{threestate1,threestate2}.  In addition,
we stress that while a range of models may prove useful for exploring
the phase transition behavior near threshold (see, for
example,~\cite{threestate1,threestate2}), only models with $W=0$
provide physically appealing characteristics far above threshold
(see, for example,~\cite{threestate3}).  Specifically, only for $W=0$
does the frequency of a perfectly synchronized set of oscillators
maintain a nonzero finite value ($g$).  Below we explore in more
detail the nature of the Hopf bifurcation associated with the class
of models described by the permitted values $(U,V,W)$.

\begin{figure}
\begin{center}
\includegraphics[width=7.0 cm]{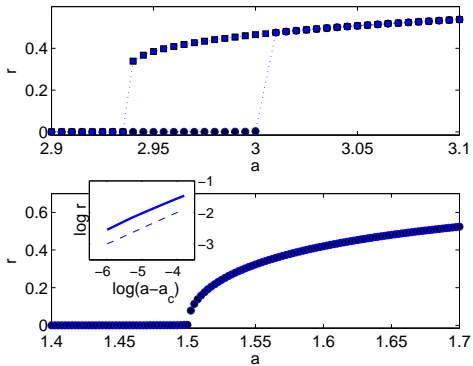}
\caption{(Color online) In a single population of globally coupled
oscillators, two physically distinct Hopf transitions can be observed
depending on the choices of $U$, $V$, and $W$.  The top panel
represents $(U,V,W)=(1,-2,0)$ and clearly shows characteristics of
a continous transition, including hysteresis.  Squares represent
solutions starting from ordered (mostly synchronized) initial
conditions, while circles represent solutions starting from disordered
(random) initial conditions.  The bottom panel represents
$(U,V,W)=(2,-1,0)$ and displays a continous transition with
critical exponent $\beta$ given by the classical value 1/2.  The
inset shows a log-log plot near the critical point.  For comparison,
a dashed line with a slope of 1/2 is shown along with the order
parameter curve (solid line) to verify this scaling relation.}
\label{SingPopSubSuperfig}
\end{center}
\end{figure}

In addition to the single population case, we consider globally
coupled arrays of oscillators that can have one of $\mathcal{N}\leq N$
different transition rate parameters,
$g=\gamma_u$, $u=1,\ldots,\mathcal{N}$. 
As detailed in~\cite{threestate3}, the probability
vector is now $3\mathcal{N}$-dimensional, $\boldsymbol{P}(t)
=(P_{1,\gamma_{1}}~P_{2,\gamma_{1}}~P_{3,\gamma_{1}}~\cdots~P_{1, \gamma_{\mathcal{N}}}~P_{2,\gamma_{\mathcal{N}}}~P_{3,\gamma_{\mathcal{N}}})^T$,
and the added subscript on the components of $\boldsymbol{P}(t)$
keeps track of the transition rate parameter.
Explicitly, the component $P_{i,\gamma_u}$ is the probability that a unit
with transition rate parameter $g=\gamma_u$ is in state $i$. 
The evolution of the probability vector is given by the set of
coupled nonlinear differential
equations $\partial \boldsymbol{P}(t)/\partial t =
M_{\mathcal{N}}[\boldsymbol{P}(t)]\boldsymbol{P}(t)$, with
\begin{equation}
M_{\mathcal{N}}[\boldsymbol{P}(t)] =
\begin{pmatrix} \mathcal{M}_{\gamma_1} & 0 &
\ldots & 0 \\ 0 & \mathcal{M}_{\gamma_2 } & \ldots & 0\\ : &: & : & : \\
0 & \ldots & 0 & \mathcal{M}_{\gamma_{\mathcal{N}}} \end{pmatrix}.  
\label{MmatmnNUnits}
\end{equation}
Here 
\begin{equation}
\mathcal{M}_{\gamma_u} = \begin{pmatrix} -g_{1}(\gamma_{u})
& 0 & g_{3}(\gamma_{u}) \\ g_{1}(\gamma_{u}) & -g_{2}(\gamma_{u})
& 0 \\ 0 & g_{2}(\gamma_{u}) & -g_{3}(\gamma_{u}) \end{pmatrix},  
\label{MBlocks}
\end{equation}
and
\begin{equation}
\begin{aligned}
g_{i}(\gamma_{u}) &= \gamma_{u} \exp\bigg[a\sum_{k=1}^{\mathcal{N}}
\varphi(\gamma_k) \\ & \times \left(U P_{i+1,\gamma_k}+ V P_{i-1,\gamma_k}
+ W P_{i,\gamma_k}\right)\bigg].
\end{aligned}
\label{gmuNunits}
\end{equation}
The function $\varphi(\gamma_k)$ is the fraction
of units which have a transition rate parameter $g=\gamma_k$. 

Because it closely appeals to physical intuition~\cite{threestate3}
for oscillators far above threshold, we limit ourselves to the case
$(U,V,W)=(1,-1,0)$ for dichotomously disordered oscillators.  We further
limit our focus here to uniform distributions
$\varphi(\gamma_k) = 1/{\mathcal{N}}$, but note that relaxing
this constraint has been shown to preserve the qualitative features of
the model~\cite{threestate3}.  For uniform distributions and
$(U,V,W)=(1,-1,0)$, probability normalization again allows us to
reduce this to a system of $2\mathcal{N}$ coupled ordinary differential
equations.  We can then linearize about the disordered state
$\boldsymbol{P}(t) = (1/3~1/3~\ldots ~1/3)^T$
and arrive at a $2\mathcal{N}\times 2\mathcal{N}$ Jacobian parameterized
by a collection of $\mathcal{N}$ transition rate parameters
${\gamma_i}$ and a coupling strength $a$.

While it has been shown that the qualitative essence of the model
remains similar for $\mathcal{N}=2,3,4$ and even for completely
disordered populations~\cite{threestate3}, we focus here only on
the simple dichotomously disordered case, $\mathcal{N}=2$.  As
shown in~\cite{threestate3}, the four eigenvalues
$(\lambda_+,\lambda_+^*,\lambda_-,\lambda_-^*)$ of the corresponding
Jacobian are given by
\begin{equation}
\begin{aligned}
\label{eigenanalytical}
\frac{{\rm Re} \lambda_{\pm}}{\gamma_1+\gamma_2} &= \frac{1}{8} \left[a-6 \pm
\mathcal{B}(a,\mu)\cos\left(\mathcal{C}(a,\mu)\right)\right],\\
\frac{{\rm Im} \lambda_{\pm}}{\gamma_1+\gamma_2} &= \frac{1}{8}
\left[\sqrt{3}(a+2) \pm \mathcal{B}(a,\mu)\sin \left(
\mathcal{C}(a,\mu)\right)\right],
\end{aligned}
\end{equation}
where
\begin{equation}
\begin{aligned}
\mathcal{B}(a,\mu)&\equiv \sqrt{2} \left[a^4 - 6 a^2 \mu^2 + 3 \mu^4
(a^2+3)\right]^{1/4},\\
\mathcal{C}(a,\mu)&\equiv \frac{1}{2} \tan^{-1} \left(\frac{- \sqrt{3} (a^2
-(a+3) \mu^2)}{a^2+3(a-1) \mu^2}\right).
\end{aligned}
\end{equation}
and
\begin{equation}
\mu\equiv \frac{2|\gamma_1 - \gamma_2|}{(\gamma_1 + \gamma_2)}.  
\label{mu}
\end{equation}
Aside from an overall factor $(\gamma_1+\gamma_2)$, 
Eqs.~(\ref{eigenanalytical}) depend only on the relative width
variable $\mu$, and therefore the critical coupling $a_c$, that is,
the value of $a$ at which Re $\lambda_+=0$, depends only on $\mu$. 
As Re $\lambda_-$ does not vanish for any $a$, $a_c$ corresponds to
a Hopf bifurcation, and our $\mathcal{N}=2$ model exhibits
macroscopic oscillations indicative of large-scale cooperation.  We
note that $a_c$ increases with increasing $\mu$, indicating that a
stronger coupling is necessary to overcome increasingly different values
of $\gamma_1$ and $\gamma_2$.

In what follows, we make use of the synchrony order paramter $r$ to
characterize the emergence of phase synchrony~\cite{hong}.  This
parameter is defined as
\begin{equation}
\label{orderparamsync}
r=\langle R \rangle, \qquad
R \equiv \frac{1}{N} \lvert \sum_{j=1}^N e^{i \phi_j} \rvert .
\end{equation}
Here $\phi$ is a discrete phase, taken to be $2 \pi (k-1)/3 $
for state $k \in \lbrace 1,2,3 \rbrace$ at site $j$. The brackets
represent an average over time in the steady state and an
average over all independent trials.  Therefore, $r$ serves as a
measure of phase synchronization.

\section{Continuous and Discontinuous Transitions to Synchrony}
\label{TransitionSec}
In the mean field limit, the order of the phase transition to synchrony
is closely tied to the nature of the Hopf bifurcation.  Specifically,
a subcritical Hopf bifurcation corresponds to a discontinuous
(sometimes called first-order) phase transition, while a supercritical
Hopf bifurcation indicates a continuous (second-order) transition. 
As such, we place special emphasis in this manuscript on the sign of
$l_1$, the first Lyapunov coefficient, which provides information on the
nature of the Hopf bifurcation and, by extension, on the order of the
phase transition.

\begin{figure}
\begin{center}
\includegraphics[width=7.0 cm]{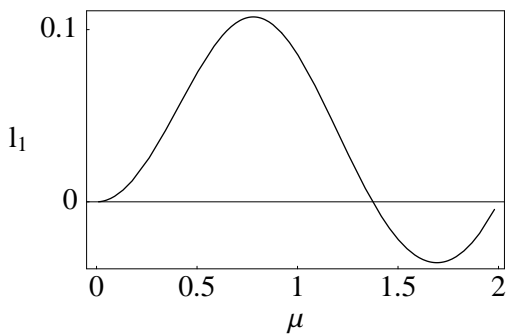}
\caption{The first Lyapunov coefficient $l_1$ is shown for Hopf
bifurcations taking place at $\epsilon^H=(a_c(\mu),\mu)$.  The
bifurcation can be either subcritical or supercritical depending on
the relative width variable.}
\label{LyapMufig}
\end{center}
\end{figure}

In general, $l_1$ can be calculated using the projection technique
given in~\cite{kuznetsov}, which relies on a multivariate Taylor
expansion of the vector field describing the dynamics in question about
an equilibrium point.  For a general $n$-dimensional dynamical system
$\dot{x}=f(x,\epsilon)$ with an equilibrium point $x=x^H$ undergoing
a Hopf bifurcation at parameter value $\epsilon=\epsilon^H$, $l_1$
is given by~\cite{kuznetsov}
\begin{equation}
\label{l1eqn}
\begin{aligned}
l_1 = \frac{1}{2\omega} & \mbox{Re} ( \langle p,C(q,q,\bar{q})
\rangle - 2 \langle p,B(q,A^{-1}B(q,\bar{q})) \rangle \\ &
+ \langle p,B \left(\bar{q},(2 i \omega I-A)^{-1}B(q,q)\right) \rangle ),
\end{aligned}
\end{equation}  
where $\langle .,. \rangle$ is the typical complex scalar product,
$I$ is the identity matrix, and $p$ and $q$ are right and left
eigenvectors of the Jacobian $A=\frac{\partial f}{\partial x} \big|_{x=x^H}$
given by
\begin{equation}
\label{qpdefeqns}
\begin{aligned}
A q &= i \omega q \\
A^T p &= - i \omega p.
\end{aligned}
\end{equation}
Furthermore, $p$ is chosen so that $\langle p,q \rangle = 1$, and
$B(u,v)$ and $C(u,v,w)$ are multilinear, $n$-dimensional vector
functions corresponding to the lowest order nonlinear coefficients in
the Taylor expansion of the vector field.  That is, 
\begin{equation}
\label{BCdefeqns}
\begin{aligned}
B(u,v) &=\sum_{j,k=1}^n \frac{\partial^2 f(\psi,\epsilon^H)}
{\partial \psi_j \partial \psi_k} \bigg|_{\psi=x^H} u_j v_k,  \\
C(u,v,w) &=\sum_{j,k,l=1}^n \frac{\partial^3 f(\psi,\epsilon^H)}
{\partial \psi_j \partial \psi_k \partial \psi_l} \bigg|_{\psi=x^H}
u_j v_k w_l,
\end{aligned}
\end{equation}
with $x^H$ indicating the equilibrium point of the vector field around
which we expand and $\epsilon^H$ the bifurcation parameter, $\epsilon$,
evaluated at the bifurcation point.  

\begin{figure}
\begin{center}
\includegraphics[width=8 cm]{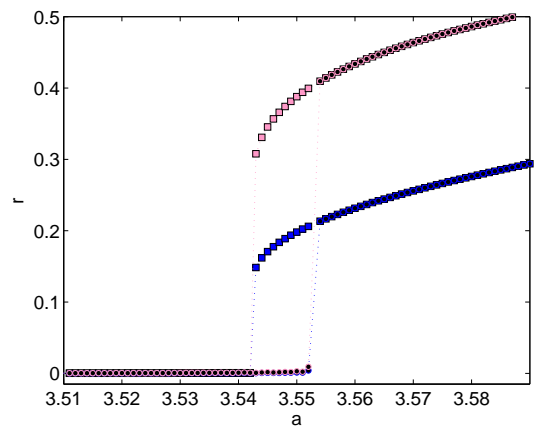}
\caption{(Color online) A subcritical Hopf bifurcation occurs for
$\mu=3/4$.  Squares represent solutions starting from ordered
(mostly synchronized) initial conditions, while circles represent
solutions starting from disordered (random) initial conditions. 
Dark (blue) points correspond to population one, $\gamma_1=2.5$, and
light (pink) points to population two, $\gamma_2=5.5$.  The transition
is clearly discontinuous as $a$ crosses $a_c \approx 3.55$.  In addition,
a region of multistability and corresponding hysteresis exists just
below threshold.}
\label{SubcriticalFig}
\end{center}
\end{figure}

\subsection{Continuous and Discontinuous Transitions in a Single
Population of Identical Oscillators}
For the case of a single population of oscillators described by
Eqs.~(\ref{gnu}) and (\ref{Mmatmn}), $l_1$ can be analytically calculated
using the technique outlined above.  Specifically, we set $g=1$
(without loss of generality) and consider the equilibrium point
$\boldsymbol{P}=(1/3,1/3)$ at $\epsilon^H \equiv a_c$ and find $q$
and $p$ to be
\begin{equation}
\label{qpBCeqn}
\begin{aligned}
q &= \left(-\frac{1}{2} + \frac{i \sqrt{3}}{2},1 \right), \\
p &= \left(\frac{i-\sqrt{3}}{3i + \sqrt{3}},\frac{2i}{3i + \sqrt{3}} 
\right), \\
\end{aligned}
\end{equation}
independent of $U$, $V$, and $W$.  Then, calculating the multivariable
functions $B(u,v)$ and $C(u,v,w)$ with Eq.~(\ref{BCdefeqns})
and using $\omega$ as defined in Eq.~(\ref{omegaeqn}) along with
Eqs.~(\ref{l1eqn}) and (\ref{qpBCeqn}), we find after simplification that
\begin{equation}
l_1=-\frac{9 \sqrt{3} (U+V-2W)}{4 \Delta_{UW}}.
\end{equation}
As a result, the nature of the Hopf bifurcation depends on the choices
$U$, $V$, and $W$.  Specifically, if we assume $U>V$, we have 
\begin{equation}
\begin{aligned}
l_1 &< 0 \mbox{ for } W< \frac{U+V}{2}; \\
l_1 &> 0 \mbox{ for } W>\frac{U+V}{2}.
\end{aligned}
\end{equation}
A similar result holds for $U<V$, but we shall here restrict ourselves to
the intuitively reasonable models positing $U \ge 0$ and $V \le 0$; that
is, the oscillators one state ahead of the one in question can only
increase (or not affect) the transition rate and those behind can only
decrease (or not affect) the transition rate.  To verify these predictions,
we show numerical solutions to the mean field equations in
Fig.~\ref{SingPopSubSuperfig}; the top panel represents an example in
the subcritical regime ($(U,V,W)=(1,-2,0)$) while the bottom panel shows
an example in the supercritical regime ($(U,V,W)=(2,-1,0)$).  A clear
distinction can be made in the neighborhood of the critical point. 
We also note that the continuous transition is characterized by the
classical mean field exponent $\beta=1/2$.

We further observe that the choice $(U,V,W)=(1,0,-1)$ leads to
$l_1 = - 27 \sqrt{3}/4 \approx - 11.69$, indicating a supercritical
Hopf bifurcation and rendering the model applicable to studies of
continuous phase transitions~\cite{threestate1,threestate2}.  With
universality in mind, we stress that any choice of parameters $(U,V,W)$
yielding a supercritical bifurcation should show similar critical behavior. 
On the other hand, the choice $(U,V,W)=(1,-1,0)$, while physically
appealing above threshold, falls at a singular point separating the
subcritical and supercritical cases ($l_1=0$).  The flexibility inherent
in the choice of coefficients $X$, $Y$, and $Z$ speaks to the richness
of our generic three-state oscillator and highlights its utility in
studying synchronization in both supercritical
(see~\cite{threestate1,threestate2}) and subcritical regimes.  We 
proceed to study the model in the presence of dichotomous disorder and
show that, for a given choice $(U,V,W)$, the level of disorder can also
alter the nature of the Hopf bifurcation and hence the order of the
phase transition.  We select the physically appealing choice $(1,-1,0)$
and, while the behavior near the critical point may depend in some sense
on this choice, we stress that our overarching goal remains unhindered. 
That is, we are able to provide an example indicating that disorder alone
can affect the nature of the transition.  The ubiquity of this
phenomenon across the entire range of models remains an open question
for future work, though we note that similar results are observed
for all parameter choices mentioned in this paper.

\begin{figure}
\begin{center}
\includegraphics[width=7.5 cm]{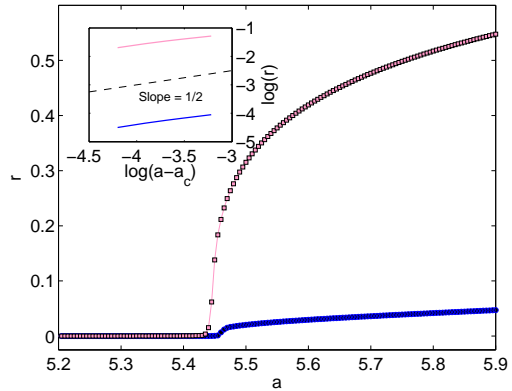}
\caption{(Color online) A supercritical Hopf bifurcation occurs
for $\mu=7/4$.  Squares represent solutions starting from an
ordered (mostly synchronized) initial condition, while circles
represent solutions starting from a disordered (random) initial
condition (repetition with other ordered and disordered
initial conditions leads to essentially the same results).  Dark (blue)
points correspond to population one,
$\gamma_1=0.25$, and light (pink) points to population
two, $\gamma_2=3.75$.  The transition is clearly continuous as $a$
crosses $a_c \approx 5.44$, and there is a noticeable absence of
hysteresis.  Dotted line is drawn to guide the eye.  The inset indicates
that, in the neighborhood of the critical point, the order parameter
follows power law behavior with the correct mean field critical
exponent ($\beta=1/2$). }
\label{SupercriticalFig}
\end{center}
\end{figure}

\subsection{Continuous and Discontinuous Transitions in a
Dichotomously Disordered Population}
Interestingly, the dichotomously disordered system corresponding to
Eqs.~(\ref{MmatmnNUnits}), (\ref{MBlocks}), and (\ref{gmuNunits})
with $\mathcal{N}=2$ and $(U,V,W) = (1,-1,0)$ can undergo either a
subcritical or supercritical bifurcation depending on the value
of $\mu$ characterizing the individual transition rates.  The transition
to synchrony occurs at a single value of the coupling $a_c(\mu)$
dependent on the relative width parameter~\cite{threestate3}.  As
such, $a$ and $\mu$ are not truly independent parameters, and we can
in principle eliminate $a$ and consider $\mu$ to be the bifurcation
parameter of interest.  Then, using the machinery of
Eqs.~(\ref{l1eqn}), (\ref{qpdefeqns}), and (\ref{BCdefeqns}), it is a
straightforward but tedious exercise to numerically evaluate the
first Lyapunov coefficient $l_1(\mu)$ corresponding to the Hopf
bifurcation occurring at $(\mu,a(\mu))$.  As shown in
Fig.~\ref{LyapMufig}, the sign of $l_1$ varies depending on the
relative width parameter (which in turn determines the critical
coupling $a_c$).  Hence, the phase transition to synchrony can appear
continuous or discontinuous depending on the relative difference between
the transition rate parameters in the two populations.        

\begin{figure}
\begin{center}
\includegraphics[width=8.0 cm]{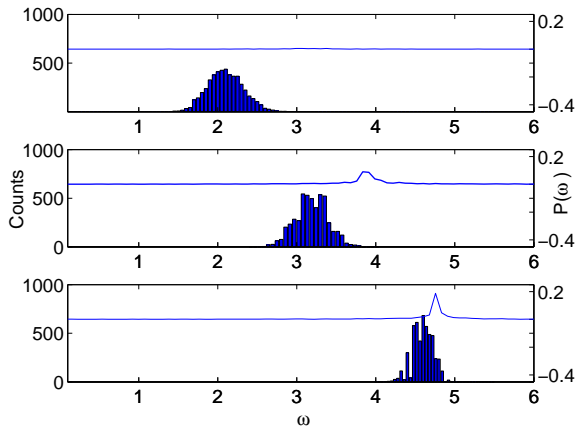}
\caption{(Color online) Each plot shows a histogram of time-averaged
frequencies (in the steady state), where the vertical axis represents
the number of units (out of $N=3500$ total units) having the
frequency $\bar{\omega}$.  The power spectrum of $P_{1,\gamma_1}$
overlays each histogram.  The top panel is below synchronization
threshold ($a=2.65$), while the middle ($a=3.05$) and lower panels
($a=3.45$) are both above threshold.} 
\label{SyncUnitsHistsSingPop}
\end{center}
\end{figure}

To verify these predictions, we solve the mean field equations
numerically in both the subcritical ($\mu=3/4$) and supercritical
($\mu=7/4$) regimes.  In the former case, we consider the case
$\gamma_1=2.5$, $\gamma_2=5.5$.  Figure~\ref{SubcriticalFig} clearly
indicates that the transition to synchrony is marked by a
discontinuous change in the order parameter $r$ as $a$ eclipses
$a_c \approx 3.55$.  In addition, a small region of marked hysteresis
appears just below threshold.  Remarkably, this indicates that a
stable disordered solution coexists with a stable, synchronized
solution (the stable limit cycle) just before threshold.  

By contrast, the case $\mu=7/4$ corresponds to a supercritical
Hopf bifurcation reminiscent of a continuous phase transition.  As
shown in Fig~\ref{SupercriticalFig}, the transition is characterized
by a continously increasing order parameter; no hysteresis is
evident.  We note also that the order parameter displays a power law
increase near the onset of the bifurcation marked by the mean field
critical exponent $\beta=1/2$.  This is expected both from the Hopf
bifurcation theorem, which prescribes the $(a-a_c)^{1/2}$ dependence
of the limit cycle radius (closely related to $r$, the order parameter)
near the onset of synchrony, and also because of the analogy with
phase transitions in an infinite-dimensional space
(see~\cite{threestate1,threestate2}).  

Interestingly, these results indicate that the degree of spatial disorder
may fundamentally alter the nature of the phase transition to synchrony. 
In both the subcritical and supercritical cases, synchronization is marked
by the destabilization of the non-synchronous state at a single value
of $a_c$, giving rise to emergent oscillations in a macroscopic variable
[for example, $P_{1,\gamma_i}(t)$].  Hence, both cases retain the
qualitative features of synchronization in disordered populations
discussed in our previous work~\cite{threestate3}; however, the
details of the onset of such cooperation distinguish the two cases.  

\begin{figure}
\begin{center}
\includegraphics[width=8.0 cm]{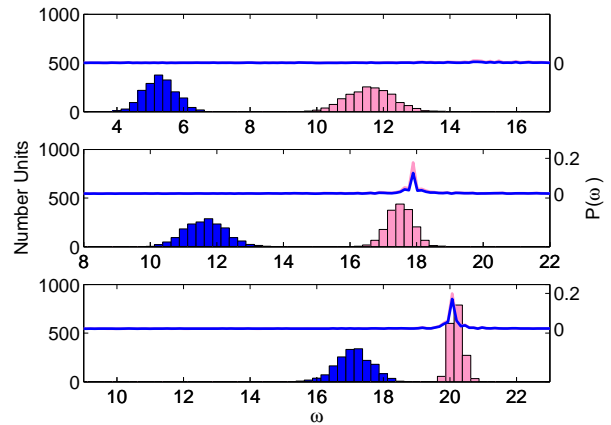}
\caption{(Color online) Each plot shows a histogram of time-averaged
frequencies (in the steady state), where the vertical axis represents
the number of units (out of $N=3500$ total units) having the
frequency $\bar{\omega}$.  Population one, characterized by
$\gamma_1=2.5$, is represented by the dark (blue) histogram, while
population two, characterized by $\gamma_2=5.5$, is represented by
the light (pink) one.  Power spectra of $P_{1,\gamma_1}$ (dark, blue)
and $P_{1,\gamma_2}$ (light, pink) overlay the histograms.  The top
panel is below synchronization threshold ($a=3.20$), while the
middle ($a=3.60$) and lower panels ($a=3.86$) are just above threshold. 
For aesthetic purposes, the horizontal range is relatively shifted in
the three panels.}
\label{SyncUnitsHists}
\end{center}
\end{figure}

\section{Microscopic Underpinnings of Synchronization}
\label{underpinnings}
Having detailed two distinct mechanisms by which synchronization might
arise, we now explore in detail the microscopic subtleties underlying
synchronization above threshold.  As detailed
in~\cite{threestate1,threestate2,threestate3} and mentioned above,
synchronization occurs in the mean field limit via the destabilization
of a nonsynchronous fixed point.  Specifically, a single pair of complex
conjugate eigenvalues corresponding to the linearized fixed point cross
the imaginary axis at $a_c$, giving rise to stable oscillations in
the macroscopic variables characterizing the system [in our case,
the components of $\boldsymbol{P}(t)$].  While the onset of this behavior
is dependent on the choice $(U,V,W)$ and also the magnitude of disorder
within the system (Sec.~\ref{TransitionSec}), the qualitative features
of the synchronized state remain identical above threshold in
both the subcritical and supercritical case.  Hence, we limit our
attention to several illustrative cases, but note that our results
hold also for the supercritical case (and in fact then entire range
of $\mu$).  Specifically, in what follows, we take $(U,V,W)=(1,-1,0)$
and consider a single population as well as a dichotomously
disordered population with $\mu=3/4$.

\begin{figure}
\begin{center}
\includegraphics[width=7.5 cm]{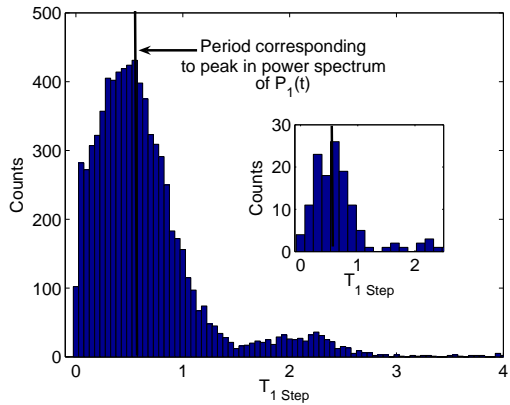}
\caption{(Color online) The central figure shows a histogram of waiting
times $T_{1 step}$ for a sampling of $N=60$ units once the steady state
has been reached.  The vertical black line indicates the waiting
time corresponding to the peak in the power spectrum
of $P_1(t)$ (that is, the waiting time corresponding to the frequency
of the macroscopic oscillations).  The inset shows a similar histogram
for a single unit.  $a=3.05$ (above threshold), $\gamma=1$ in all plots.}
\label{SyncStepsHistsSingUnit}
\end{center}
\end{figure}

\begin{figure}
\begin{center}
\includegraphics[width=7.5 cm]{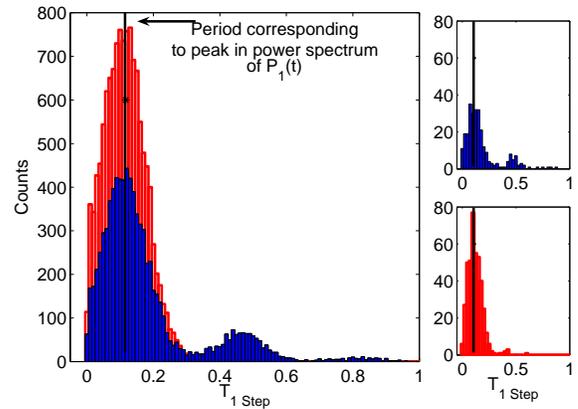}
\caption{(Color online) The central figure shows a histogram of waiting
times $T_{1 step}$ for a sampling of $N=30$ units from each population
once the steady state has been reached.  Dark (blue) represents
population one, while light (red) represents population 2.  The
vertical black line indicates the waiting time corresponding to the
peak in the power spectrum of $P_1(t)$ (that is, the waiting time
corresponding to the frequency of the macroscopic oscillations). 
The insets show similar histograms for single units; the top histogram
is for a unit from population one, the bottom from population two. 
$a=3.60$ (above threshold), $\gamma_1=2.5$, and $\gamma_2=5.5$ in all
plots.}
\label{SyncStepsHists}
\end{center}
\end{figure}

In particular, the threshold $a_c$ is marked by the onset of coherent
temporal oscillations in the components of $\boldsymbol{P}(t)$. 
We characterize the microscopic underpinnings of these oscillations by
considering $\bar{\omega_i}$, the time-averaged frequency of oscillator
$i$ in the steady state.  We perform simulations on globally coupled
lattices of $N=3500$ units of a single population with $\gamma=1$ and
also of a dichotomously disordered population with $\gamma_1=2.5$
and $\gamma_2=5.5$.  As shown in Figs.~\ref{SyncUnitsHistsSingPop}
and~\ref{SyncUnitsHists}, the distribution of frequencies $\bar{\omega_i}$
clusters around the values prescribed by $\gamma$ (or $\gamma_1$
and $\gamma_2$ for populations one and two, respectively) far below
threshold (top panels).  Specifically, for a deterministic oscillator
with transition rate $\gamma$, $\bar{\omega_i}$ is given by
$2 \pi \gamma/3$; when $\gamma=1$ (or $\gamma_1=2.5$ and $\gamma_2=5.5$),
this gives the central peak of the histogram for the relevant population.
We compare these histograms with the power spectra $\left(
\tilde{P}_{1,\gamma_i}(\omega)
\tilde{P}^*_{1,\gamma_i}(\omega)\right)^{1/2}$, where
$\tilde{P}_{1,\gamma_i}(\omega)$ is the Fourier transform of
$P_{1,\gamma_i}(t)$.
As threshold is eclipsed (middle panels), a peak arises in the power
spectrum of the macroscopic variables $P_{1,\gamma_i}$, though the
frequency of this peak does not correspond with the individual
$\bar{\omega_i}$'s of oscillators constituting the population.  In
the dichotomous case, this peak only roughly corresponds with the time
averaged frequencies from population two and completely exceeds even
the maximum $\bar{\omega_{i}}$ characterizing population one.  As
$a$ is further increased, the descrepency between the time-averaged
frequency histograms and the macroscopic oscillation frequencies
decreases.  In addition, in the disordered case, the histograms for
the two populations become increasingly narrow and closer to one
another (bottom panel).  We note that as $a$ becomes tremendously
large, the histograms become extremely narrow and begin to overlap
at a frequency determined by the frequency of the macroscopic
oscillations, as expected (indicative of perfect synchronization). 
Nonetheless, the behavior for finite, intermediate $a$ is rather
counterintuitive and points to a rich microscopic dynamics underlying
the cooperative behavior.      

\begin{figure}
\begin{center}
\includegraphics[width=7.5 cm]{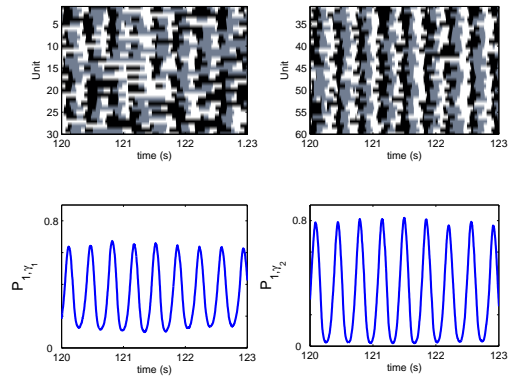}
\caption{(Color online) The top panels show the evolution of
a representative sub-system ($N=30$ units of each population).  
Dark, medium, and light represent states 1,2, and 3, respectively. 
The bottom panels show the macroscopic variable $P_{1,\gamma_i}$
for each population.  The left panels correspond to population
one ($\gamma_1=2.5$), while the right panels correspond to population
two ($\gamma_2=5.5$).  $a=3.60$ (above threshold) for all plots.}
\label{SyncPartVisualize}
\end{center}
\end{figure}

To further explore these trends, we consider the stochastic variable
$T_{1~step}$, the waiting time in a single state for an individual
oscillator.  $T_{1~step}$ represents the time the oscillator spends
in a single state $i$ before transitioning to the subsequent state
$i+1$.  For computational efficiency, we record $T_{1~step}$ for
a representative subpopulation of $60$ units ($30$ units from each
population, one and two, in the disordered case). 
Figures~\ref{SyncStepsHistsSingUnit} and~\ref{SyncStepsHists} show
histograms of the variable $T_{1~step}$ taken over this representative
subpopulation once steady state was reached.  Clearly, all relevant
subpopulations consist of oscillators whose steps most often correspond
to the frequency of the macroscopic oscillation (shown by the solid
vertical line).  That is, the peak of the histograms occur at a
value $T$ comensurate with the frequency peak in the power spectrum of
the components of $\boldsymbol{P}(t)$.  However,
Fig.~\ref{SyncStepsHistsSingUnit} shows that the distribution of
$T_{1~step}$ is bimodal, with a significant peak occuring at
$T_{1~step} \approx 2.2$ which downward biases the time-averaged
frequencies $\omega_i$ of individual units.  We note that as coupling
$a$ increases significantly above threshold, the distribution becomes
unimodal with a peak at $T_{1~step}$ corresponding to the frequency
of macroscopic oscillation.  In the disordered case, only population one,
characterized by significantly lower time-averaged $\bar{\omega}$'s,
shows a bimodal distribution with a significant peak at
$T_{1~step} \approx 0.45$.  In fact, these long waiting times, while
not the dominant macroscopic behavior, pervade the microscopic dynamics
in such a way that the time-averaged frequencies become downward
biased and no longer accurately represent the macroscopic dynamics. 
Interestingly, population two has become sufficiently synchronized that
the second peak is effectively nonexistent, and thus the frequencies
overlap more closely with the macroscopic ``mean field" frequency. 
The right insets of Figs.~\ref{SyncStepsHistsSingUnit}
and~\ref{SyncStepsHists} show histograms for single units chosen from
the populations (or subpopulations).  Again, the unit chosen from the
single population case shows a bimodal distribution with significant
``anamolous" peaks near $T_{1~step} \approx 2.2$.  In the disordered case,
the unit from population one shows a bimodal waiting time distribution
characterized by occasional waiting times in the neighborhood of
$T \approx 0.45$ in addition to those corresponding to the macroscopic
oscillations.  Finally, in Fig.~\ref{SyncPartVisualize} we show the
time evolution of the subpopulations along with the macroscopic
variable $P_{1,\gamma_i}$ for each population.  At any given 
time, the majority of oscillators in each population is synchronized,
leading to the smooth oscillations of the macroscopic variable.  However,
isolated single units are prone to long waiting times, particularly
in the less synchronized population (population one, left panel, in
this example).  These anamolously long waiting times, which serve to
bias the time averaged frequencies $\bar{\omega_i}$ of each individual
unit, nevertheless do not substantially disrupt the macroscopic
oscillations, largely because the occurance of coincident long waits
is fairly uncommon.  That is, the long waiting times do not appear
in any significantly correlated way among individual constituents
of the population. 

\section{Discussion}
\label{discussion}
We have shown that a class of simple, discrete models of stochastic
phase coupled oscillators can undergo either a subcritical or
supercritical bifurcation to macroscopic synchrony, depending on the
chosen form of the microscopic coupling.  As such, the different
instances of the model can be used to study either continuous phase
transitions~\cite{threestate1,threestate2} or discontinuous
transitions exhibiting hysteresis, a characterstic seen in detailed
theoretical models of, e.g., coupled Josephson
junctions~\cite{filatrella07} but only observed in significantly more
complex coupled oscillator
models~\cite{daido,gianuzzi07,filatrella07,tanaka,acebron,choi,pazo,hong0}. 
We stress that universality suggests that all models in this class
exhibiting continous phase transitions should show similar behavior near
the critical point, and this served as the basis of our eariler
studies~\cite{threestate1,threestate2}.  Nevertheless, it is remarkable
that minor modifications in microscopic coupling can alter the
nature of the bifurcation in such a fundamental way.  

In addition, we have shown that in dichotomously disordered populations,
both subcritical and supercritical Hopf bifurcations can occur, and
the distinction is completely determined by the relative width
$\mu$ characterizing the transition rate disorder between the two
populations.  While the qualitative features of the transitions within
each class (subcritical and supercritical) appear identical, the
distinction between classes points to fundamentally different mechanisms
underlying the initial emergence of phase synchronization.  In particular,
it is striking that the level of disorder within a population, as
measured by $\mu$, can significantly alter the behavior near the
critical point (though we stress that behavior even moderately above
threshold is qualitatively indistinguishable).

Finally, we have studied the microscopic basis of phase synchronization
above threshold.  It is initially counterintuitive that phase
synchronization, defined in terms of the Hopf bifurcation and
temporal oscillations in the macroscopic variable $\boldsymbol{P}(t)$
(and measured in the order parameter $r$), is not contingent upon the
existence of overlapping distributions of $\bar{\omega_i}$.  That is,
our results regarding the discrete oscillator model highlight the
complexity of microscopic dynamics underlying macroscopic cooperation
and point to a potentially misleading subtlety.  Whereas phase
synchronization is often considered a stronger condition than
frequency entrainment -- defined using an order parameter built upon
the notion that a fraction of units display identical time-averaged
frequencies in the oscillator population -- we here report subtle
microscopic features which distinguish the two without establishing a
clear hierarchy.  For example, Hong et al.~\cite{hong} show that for
disordered populations of Kuramoto oscillators, the lower critical
dimension for frequency entrainment is lower than that for phase
synchronization in locally coupled oscillators, indicating the relative
ease with which frequency entrainment is achieved.  They note that the
two transitions coincide in the case of globally coupled units. 
Contrast that with our dichotomously disordered population, for which
phase synchronization occurs without any overlap in the frequency
distributions: that is, no oscillator from population one has the
same frequency as any oscillator from population two.  While a direct
comparison is not plausible owing to the specific differences between
models and order parameters, we stress that any order parameter related
to time-averaged measurements of frequencies would, for our model,
be misleading and provide potentially counterintuitive results. 
The emergence of a nonzero $r$, which measures phase synchronization,
corresponds with the loss of stability of the asynchronous fixed point
(the Hopf bifurcation).  This does not guarantee similar distributions
of time-averaged frequencies in the two populations; in fact, we can
readily see that synchronization occurs while the frequency distributions
are entirely distinct.  Furthermore, the frequency of the macroscopic
oscillations of the mean field does not always coincide with the
time-averaged frequencies of the oscillators constituting the population
(or any subpopulation).  Only when coupling is sufficiently large to
substantially reduce the anamolously long waiting times which bias
$\bar{\omega_i}$ will the frequency distributions begin to overlap
one another and coincide with the frequency of the mean field
oscillations.  Because these long waiting times appear more readily
in the population with the smaller $\gamma_i$, the time-averaged
frequencies of the two populations are disproportionately affected,
meaning that the populations will appear to behave quite differently
in terms of average frequency.  This in fact underlies the stark
differences in the degree of synchronization between two populations
as measured by the order parameter $r$, and provides an intuitive
description capable of explaining this discrepancy.  Our previous
results show that completely disordered populations show qualitative
similarities with the dichotomously disordered case~\cite{threestate3};
hence, we are led to cautiously speculate that wholly disordered
populations are also characterized by waiting times $T_{1~step}$
distributed with long tails, and hence time-averaged frequencies
become downwardly biased, meaning that the order parameter for
frequency entrainment, in the typical sense, will not accurately
reflect the macroscopic cooperation.  Further studies along these
lines are currently in progress.

Finally, the results of this work raise the following question: how
dependent is the above phenomenon on the choice of a discrete phase
model?  Would similarly counterintuitve results arise in continous
phase oscillators?  In fact, a recent study by Rosenblum and
Pikovsky~\cite{rosenblum07} suggests that a similar (though not
identical) state of partial synchronization arises in continous oscillators
coupled in a highly nonlinear fashion.  Specifically, they find that
in globally coupled oscillators, phases exist in which certain
subpopulations are characterized by time-averaged frequencies which
are not commensurate with the oscillations of the mean field, that is,
they are not locked with the macroscopic oscillations induced in
the population.  While once again the differences between the models
make direct comparison difficult, it is nevertheless clear that
measurements of time-averaged frequencies provide potentially
counterintuitive results, even in globally coupled arrays.  In the case
of our stochastic discrete oscillators, the behvavior is quite
transparent once viewed in terms of $T_{1~step}$, though it is not
clear whether a similar mechanism underlies the phenomenon in the
continous phase model.  Uncovering the relationship between the
superthreshold phase in our model and that in the continous oscillator
model of~\cite{rosenblum07} remains an open question, but even the
superficial similarities between the results motivate continued efforts
along these lines.

\section*{Acknowledgments}
This work was partially supported by the National Science Foundation
under Grant No. PHY-0354937.

\end{document}